\documentclass[aps,prl,reprint,superscriptaddress,showkeys]{revtex4-2}

\usepackage{graphicx}
\usepackage{xcolor}
\usepackage{amsmath}
\usepackage{amssymb}
\usepackage{CJK}

\begin{document}


\title{Ensemble equivalence and asymptotic equipartition property in information theory}


\author{Meizhu Li}
\affiliation{School of Computer Science and Communication Engineering, Jiangsu University, Zhenjiang, Jiangsu, China}



\author{Qi Zhang}
\email[]{qi.zhang@just.edu.cn}
\affiliation{School of Science, Jiangsu University of Science and Technology, Zhenjiang, Jiangsu, China}
\affiliation{Lorentz Institute for Theoretical Physics, Leiden University, PO Box 9504, 2300 RA Leiden, The Netherlands}


\date{\today}

\begin{abstract}
There is a consensus in science that information theory and statistical physics have a close relationship but the literary proofs of the equivalence between most of the conceptions in the two disciplines are still missing. 
In this work, according to the statistical ensembles' description of the information sequences that are generated by the i.i.d. single variable and multivariate information source, the relationship between the ensemble equivalence and asymptotic equipartition property is established. 
We find that the description of information sequences in classical information theory is a special case of the canonical ensemble description of information sequences. They are both the maximum entropy approximation of the real signal generation. Vice versa, the conjugate microcanonical ensemble description of the information sequences under hard constraints satisfies the condition in real signal generation exactly. Thus, the microcanoincal ensemble description is closer to the real signal generation than the conjugate canonical ensemble, but the ensemble equivalence between the microcanonical and canonical ensemble in the thermodynamic limit guarantees the effectiveness of classical information theory, i.e., the asymptotic equipartition property in information theory is an isotope of ensemble equivalence from statistical physics. 
\end{abstract}


\maketitle



\section{Introduction}


Information theory was built by Shannon in 1948 based on the publication of his fundamental article, which introduced the theoretical limits of information transition and compression in communication systems~\cite{shannon1948mathematical}. Shannon found that the smallest space needed to store the information generated by the information sources is determined by the entropy of the information source~\cite{cover1999elements}, which often serves as the guiding principle for the source coding theorem and lossy compression. Actually, this finding can be derived from the introduction of the asymptotic equipartition property.

Asymptotic equipartition property shows that in lossy compression, most of the information generated by identical independent information sources is carried by the equiprobable sequences belonging to the typical set. When the length of the information sequences goes infinite, the probability of those equiprobable sequences appearing in the process of real signal generation is decided by the probability of the information source getting different states, which can be quantified by the \textit{Shannon entropy} of the information source. Thus, the smallest space needed to store the information that is generated by the information source is decided by the {Shannon entropy} of the information sources~\cite{shannon1948mathematical}. In other words, the asymptotic equipartition property shows that when the length of the information sequences goes infinite, the limit of space to store the information carried by those information sequences is decided by the entropy of the information sources. This phenomenon can also be explained as the typicality of the sequences in the typical set, i.e., those equiprobable sequences decide the macroscopic property of all the sequences, which is familiar in statistical physics~\cite{touchette2015equivalence}.   

The building of information theory is based on probability theory's description of information sources and sequences, which is the same as the probability description of the states in complex systems in statistical mechanics, i.e., the mathematical foundations of information theory and statistical physics are the same. That is why Shannon entropy in information theory has the same mathematical form as \textit{Gibbs entropy} in statistical physics, and the distribution of states in the energy-constrained system satisfies the maximum entropy principle~\cite{jaynes1957information}. Thus, when each information sequence is treated as a state of systems in statistical physics, the asymptotic equipartition property also can be explained as the macroscopic properties of a thermodynamic system are dominated by a group of 'equiprobable' states, and it is constrained by the behavior of the particle in the thermodynamic system~\cite{touchette2015equivalence}. 

In information theory, the efficacy of the asymptotic equipartition property necessitates considerable length of the information sequences. This condition resembles the prerequisite in statistical physics, wherein the scale of the systems ought to be sufficiently large. Ensemble equivalence is also under this condition, which is called thermodynamic limit~\cite{gibbs1902elementary}. 

Ensemble equivalence is a fundamental assumption in traditional statistical mechanics, postulates that the microcanonical and canonical ensemble descriptions become equivalent when the system is subjected to the thermodynamic limit~\cite{gibbs1902elementary}. In statistical physics, the microcanonical ensemble is used to describe the systems with hard constraints, e.g., isolated systems with fixed total energy. The canonical ensemble is used to describe the systems with soft constraints, e.g., systems with fixed temperatures. Therefore, ensemble equivalence means the typical states in the two ensembles are close to each other, and that is why the macroscopic properties of the two ensembles are the same in the thermodynamic limit, as enough big size of the system makes sure the fluctuation in the canonical ensemble will converge to a relatively small value~\cite{ellis2002nonequivalent,dionigi2021spectral,zhang2021statistical}.

When the information sequences generated by the information source are treated as states in thermodynamic systems, the equivalence between the two ensemble descriptions is analogous to the equivalence between the real signal generation in artificial communication systems and the process described by the classical information theory, as the signal generation can be modeled by the microcanonical ensemble with hard constraints exactly. The classical information theory is a maximum entropy approximation of the information sequence, which is a particular canonical ensemble description conjugated with the microcanonical ensemble. In other words, the asymptotic equipartition property in classical information theory is an isotope of the ensemble equivalence in statistical physics. Therefore, checking the relationship between the ensemble equivalence and asymptotic equipartition property will build a fundamental interaction between information theory and statistical physics. 

The rest of this paper is organized as follows: in section 2, the statistical ensembles are used to describe information sequences generated by the classical i.i.d, information and reformulate the classical information theory by ensemble theory. We prove that the classical information theory is a special case of the canonical ensemble description of information sequence with soft constraints. In section 3, the statistical ensembles are used to describe the information sequences generated by the independent multivariate information source. We also prove that when the information sequence is under ensemble equivalence, the asymptotic equipartition property is postulated even when the information sources are multivariate but independent. Conclusions and discussion are given in section 5. 

\section{Ensemble description of classical information theory}
Information sequences generated by the i.i.d. information sources resemble the one-dimensional Ising model without the interior interactions. Thus, the reformulation of the classical information theory is based on the ensemble description of the information sequences, both for the single variable information source and multivariate-independent information sources. 

Before the reformulation, we will introduce the typical signal-generating description in Shannon's information theory. 
In classical information theory, there is a basic assumption that each unit in the sequence is independent, as the information source is independent and identical. The probability for each unit to get different values is the same~\cite{albert2002statistical}. 
For example, in the single binary variable information source $x$, the probability of the sequence $A=[a_1,a_2,\cdots,a_i,\cdots,a_n]$ with length $n$, and $t$ units in it equal to $1$ is 
 \begin{equation}
	 P(A|t)=p^{t}(1-p)^{n-t},
 \end{equation}
where $p$ is the probability of the information source $x$ to have value of 1, and $t$ is the total number of units with value $1$ in the sequence. 

When the length $n$ of the sequence $A$ goes to infinite, and the probability $p$ of each unit to get value $1$ fixed, we can find the average number of the unit in the sequence $A$ to has value one equal to $\langle t\rangle=t^*=n\times p$. It is a manifestation of the large number law. This result can also be explained as the probability that information source $x$ having value one will be manifested as the proportion of the total number of units in the information sequences with unit value 1 to the length of the information sequence. Obviously, the proportion will converge to the probability $p$ when the length of the information sequence goes to infinite. 

As each unit to have value one in the information sequence is equiprobable, the rescaled logarithm of the probability of the information sequence ${A}$ is 
\begin{equation}
	\frac{1}{n}\ln P(A|p)=p\ln p+(1-p)\ln(1-p)
\end{equation}
When the length of the information sequence goes to infinite, and the probability converges to ($p=t^*/n$), the rescaled logarithm probability is the average limit to store the information generated by the information source, which is contained in the typical set. According to the large number law, the limit of the rescaled probability is equal to
\begin{equation}
	\lim_{n\rightarrow\infty}\frac{1}{n}\ln P(A|t^*)=\frac{t^*}{n}\ln \frac{t^*}{n}+\frac{n-t^*}{n}\ln\frac{n-t^*}{n},
\end{equation}
which is the minus of the Shannon entropy of the information source $x$, and it is equal to 
\begin{equation}
s(x)=-\frac{t^*}{n}\ln \frac{t^*}{n}-\frac{n-t^*}{n}\ln\frac{n-t^*}{n}.
\end{equation}
This finding has connected the limit of information storage with the uncertainty of the information source, and it also can be extended to the understanding of the definition of the typical set $T_{\epsilon}$, which is the set of a group of equiprobable sequences that carries almost all the information. The definition of the typical set $T_{\epsilon}$ can be explained as it is the collection of all the sequences that satisfy the following condition
 \begin{equation}
	 T_{\epsilon}=\{A|e^{-n(s({x})-\epsilon)}\leq P(A)\leq e^{-n(s({x})-\epsilon)}\}.
 \end{equation}
Then we can find that the small space to store sequences in the typical set is equal to
\begin{equation}
	\begin{aligned}
		\ln|T_{\epsilon=0}|&=n\times s(x)\\
		&=n\ln n-t^*\ln{t^*}-(n-t^*)\ln{(n-t^*)}
	\end{aligned}
\end{equation}
The result shows one of the main conclusions in Shannon's information theory that the space needed to store the information generated by the information source is decided by the uncertainty of the information source. 

As we already mentioned before, to find the relationship between the asymptotic equipartition property and ensemble equivalence, we need to find the same object that can be described by information theory and statistical physics. Therefore, in this section, the statistical ensembles will be used to describe the information sequences generated by the identical-independent distributed information source, both the microcanonical ensemble and the canonical ensemble.

\subsection{Microcanonical description of the classical information sequence}
From statistical physics' view, the information sequences $\{A\}$ generated by the binary i.i.d information source $x$ with length $n$ have $2^n$ combinations. The probability distribution of those information sequences decides the limit of space to store those information sequences. When the binary information source $x$ to generate state 1 or 0 has the same probability, each information sequence $A$ has the same probability, and the number of configurations of the information sequences decides its value. When the probability of the information source having different states is different, then estimating the probability distribution of those information sequences with different probability (i.e., information sequences under 'hard' constraints) needs the microcanonical ensemble.

The probability of these sequences with $t^*$ units with value $1$ can be obtained by the microcanonical ensemble description as 
 \begin{equation}
	 P_{\textrm{mic}}(A|t^*)=1/\binom{n}{t^*}.
 \end{equation}
The combination $\Omega_{\textrm{mic}}=\binom{n}{t^*}$ is the number of configurations of the information squences $\{A\}$ with hard constraints $t^*$.

According to the definition of asymptotic equipartition property, all the sequences in the microcanonical ensemble $\mathbf{A}_{\textrm{mic}}$ belong to the typical set $T_{\textrm{mic}}$ of it, as they all have the same probability. Therefore, the smallest space needed to store the information carried by those sequences equals $\ln|T_{\textrm{mic}}|$, and it is the Boltzmann entropy of the microcanonical ensemble 
 \begin{equation}
	\begin{aligned}
		\ln|T_{\textrm{mic}}|&=\ln\binom{n}{t^*}=\ln\Omega_{\textrm{mic}}=S_\textrm{mic}\\
		&=\ln\frac{n!}{t^*!(n-t^*)!}\\
		&\rightarrow n\ln n-t^*\ln t^*-(n-t^*)\ln(n-t^*).\\
	\end{aligned}
 \end{equation}
 where $\Omega_{\textrm{mic}}$ is the total number of sequences under this hard constraints $t^*$. 
 
 This result shows us that if the total energy can be used by the information source $x$ to generate the signal is fixed as $t^*$, and the length of the information sequences $n$ goes to infinite, the signal-generating process that is described by the microcanonical ensemble is closer to the real process. 
The smallest space needed to store the information carried by those information sequences is equal to the Boltzmann entropy of the microcanonical ensemble.

 \subsection{Canonical description of the classical information sequences}

In the microcanonical ensemble description of the information sequences, the number of units with value 1 is fixed as $t^*$(under length $n$). It means the information source can use the energy to generate a signal with value 1 is fixed as $t^*$. However, the 'hard' constraints are difficult to come true. A more flexible way to describe those information sequences is using the conjugate canonical ensemble with soft constraints. 

According to the definition of system with local constraints in~\cite{zhang2022strong}, the information sequence $A$ generated by the i.i.d. information source $x$ under global constraints $t=\sum_{i=1}^n a_i$. The 'soft' constraint is the average value of the total number of units with value 1 in the information sequence $A$ equals the hard constraints $t^*$, which can be formulated as $\langle t\rangle=t^*$. 

The maximum likelihood parameter $\theta^*$ will realize this condition and maximum the Shannon entropy of the canonical ensemble, and we can define the \emph{Hamiltonian} of the canonical ensemble described information sequence as 
\begin{equation}
	H(A)=t\cdot\theta^*.
\end{equation}
For the binary information source, the partition function is equal to 
\begin{equation}
	Z(\theta^*)=(1+e^{-\theta^*})^n.	
\end{equation}
Probability of the sequence with global constraint $t$ is 
  \begin{equation}\label{eq:p_t_*_1}
	  P_{\textrm{can}}(A|(\theta^*,t))=\frac{e^{-t\theta^*}}{(1+e^{-\theta^*})^n}.
  \end{equation}
 The soft constraints require the average value of global constraints in the canonical ensemble equals to $t^*$ as
 \begin{equation}
	 \langle t\rangle_{\textrm{can}}=\sum_{A\in\mathbf{A}_{\textrm{can}}}t(A)P_{\textrm{can}}(A|\theta^*,t)=t^*,
 \end{equation}
 where $\mathbf{A}_{\textrm{can}}$ represents all the information sequences that are generated by the information sources (all the information sequences described by the canonical ensemble). Therefore, we can find that 
the probability of unit $a_i$ to have value 1 in the canonical ensemble is equal to
 \begin{equation}
	 p=\frac{e^{-\theta^*}}{1+e^{-\theta^*}}=\frac{t^*}{n},
 \end{equation}
 as $e^{-\theta^*}=\frac{t^*}{n-t^*}$.

 Then, we can find that the Shannon entropy of the information source under canonical ensemble description is equal to 
 \begin{equation}
	 s_{\textrm{can}}(x)=-[\frac{t^*}{n}\ln \frac{t^*}{n}+\frac{n-t^*}{n}\ln\frac{n-t^*}{n}].
\end{equation}
 Hence, the typical set of sequences in the classical information theory can be obtained from the AEP as 
  \begin{equation}
	 \begin{aligned}
		 \lim_{n\rightarrow\infty}\frac{1}{n}\ln P_{\textrm{can}}(A|\theta^*,t)&=\frac{1}{n}\sum_{i=1}^n\ln p(a_i)\\
		 &\rightarrow E\ln p(x)\\
		 &=-s_{\textrm{can}}(x),	
	 \end{aligned}
  \end{equation}
The sequences in the typical set should satisfy the following condition 
 \begin{equation}
	 T^{\epsilon}_{\textrm{can}}=\{A|e^{-ns_{\textrm{can}}(x)-\epsilon}\leq P(A)\leq e^{-ns_{\textrm{can}}(x)+\epsilon}\}.
 \end{equation}
 According to the relationship shown in Eq.(\ref{eq:p_t_*_1}), sequences belong to the typical set also can be 
 represented by the Shannon entropy of the canonical ensemble. As the sum of $n$ rescaled entropy $s_{\textrm{can}}(x)$ is equal to 
 \begin{equation}
	 \begin{aligned}
	 n\times s_{\textrm{can}}(x)&=\frac{{t^*}^{t^*}(n-t^*)^{n-t^*}}{n^n}\\
	 &=\ln P_{\textrm{can}}(A|(\theta^*,t^*))	\\	
	 &=S_{\textrm{can}}
	 \end{aligned}
 \end{equation}
 Thus, the space to store the information carried by the canonical ensemble described information sequence equals
 \begin{equation}
	\begin{aligned}
		\ln|T^{\epsilon}_{\textrm{can}}|&=S_{\textrm{can}}=n\times s(x)\\
		&=-t^*\ln\frac{t^*}{n}-(n-t^*)\ln\frac{n-t^*}{n}\\
		&=n\ln n-t^*\ln t^*-(n-t^*)\ln(n-t^*).
	\end{aligned}
 \end{equation}

 The result of the canonical ensemble description shown above is equivalent to the classical description of the information sources. It proves that classical information theory is a canonical ensemble description when the interactions in the information sequences are homogeneous.

 \subsection{Ensemble equivalence and asymptotic equipartition property in classical information theory}
We can find that the canonical ensemble description shown above is equivalent to the classical information theory when the i.i.d. assumption holds. If the length of the information sequence goes to infinite $n\rightarrow\infty$, the canonical ensemble description of the information sequences is also equivalent to the microcanonical ensemble (the Stirling approximation), but we still need a quantified measurement of the difference between the microcanonical and canonical ensemble description of the information sequence, when the information source is i.i.d.

The macroscopic difference between the two ensemble descriptions of the information sequences is the difference in the space to store the information carried by the sequences. This difference is equal to the difference between the entropy of the two ensembles  
 \begin{equation}
	\ln|T^{\epsilon}_{\textrm{can}}|-\ln|T^{\epsilon}_{\textrm{mic}}|=S_{\textrm{can}}-S_{\textrm{mic}}.
 \end{equation}

 As we know, the states in the microcanonical ensemble are the subset of the state in the conjugated canonical ensemble, $\mathbf{A}_{\textrm{mic}}\in \mathbf{A}_{\textrm{can}}$, so the difference on the space to store the information carried by different ensemble described information sequences can be formulated as the relative entropy between the probability distribution of the microcanonical ensemble and the conjugated canonical ensemble as
 \begin{equation}
	\begin{aligned}
		S(P_{\textrm{mic}}||P_{\textrm{can}})&=\sum_{{A}\in\mathcal{A}}P_{\textrm{mic}}({A}|t^*)\ln\frac{P_{\textrm{mic}}({A}|t^*)}{P_{\textrm{can}}({A}|\theta^*)}\\
	&=S_{\textrm{can}}-S_{\textrm{mic}}\\
	&=\ln|T^{\epsilon}_{\textrm{can}}|-\ln|T^{\epsilon}_{\textrm{mic}}|\\
	&=\Delta S_{\textrm{can}}	
	\end{aligned}
 \end{equation}
It means the difference in the limit of information storage is equal to the difference in the entropy of the microcanonical ensemble and canonical ensemble. 
When the relative entropy is rescaled by the length of the information sequence $n$, the value of it equals 
 \begin{equation}
	\frac{1}{n}\Delta S_{\textrm{can}}=\frac{1}{n}[\ln\frac{n^n}{{t^*}^{t^*}(n-t^*)^{n-t^*}}-\ln\binom{n}{t^*}].
 \end{equation}
 According to the Stirling approximation, the limit of the rescaled difference $\frac{1}{n}\Delta S_{\textrm{can}}$ equals 0
 \begin{equation}
	 \lim_{n\rightarrow\infty}\frac{1}{n}\Delta S_{\textrm{can}}=\lim_{n\rightarrow\infty}\frac{1}{n}[\frac{1}{2}\ln[2\pi{t^*(1-\frac{t^*}{n})}]]=0.
 \end{equation} 
It means the canonical ensemble will converge to the microcanonical one in the thermodynamic limit. It also shows that the limit of the information storage is the same when the sequences are described by the two ensemble descriptions. 
In statistical physics, this phenomenon is called measure-level ensemble equivalence~\cite{touchette2015equivalence}. The logarithm difference $\ln|T^{\epsilon}_{\textrm{can}}|-\ln|T^{\epsilon}_{\textrm{mic}}|$ is the relative entropy, which grows like $o(n)$. Based on this assumption, the cannonical ensemble is always used to replace 
the mathematically difficult microcanonical ensemble in the theoretical research and practical application.

The microcanonical ensemble description has realized the constraints in the signal generating exactly. The classical information theory is a particular example of the canonical ensemble descriptions. The ensemble equivalence between the microcanonical and canonical ensemble descriptions under the global constraints $t^*$ indicates why the classical information theory is effective, as this ensemble equivalence allows the description in the classical information theory close to the process of signal generation when the length of the sequences goes to infinite. 

\section{Ensemble description of information sequences from multivariate i.i.d. information source }
\label{Section:row_local}
The ensemble description of the classical information sequences shows that the classical information theory is an isotope of the canonical ensemble description. The two theories are based on the same assumption: the distribution of states is based on the maximum entropy principle~\cite{jaynes1957information}. Actually, compared with the description of information sequences generated by signal variable information sources, the ensemble theory is more suitable for describing the information sequences generated by the information sources with multivariates. In this part, we will introduce how to use the statistical ensemble to describe the multivariate information source and try to prove that the equivalency between asymptotic equipartition property and ensemble equivalence is still valid. 

When the multivariate information source has $m$ units, information sequences generated by this information source are $m\times n$ matrix. This information source also has been described in the fundamental paper of Shanon in 1948~\cite{shannon1948mathematical}, which is called a "joint information source".
Thus, there are $m$ rows in the matrices ${X}$, which means there are $m$ units in the information source. The signal generating of the multivariate independent information source is under ensemble equivalence, as there is a finite number of local constraints in it, and there is no phase transition in the information sequence~\cite{zhang2022strong}. The classical information theory can still describe this signal generation. Then, according to the AEP, we can find the limit of information storage. 

The sequence generated by the information source $\vec{x}=[b_1,b_2,\cdots,b_j,\cdots,b_m]$ with $m$ independent variables is an $m\times n$ matrix ${X}$. As the $m$ units in the information source are independent of each other, the sequence $X$ can be divided into $m$ row vectors ${X}=\{\vec{r}_1;\vec{r}_2;\cdots;\vec{r}_m\}$. Each row vector $\vec{r}_j$ of the matrix ${X}$ has $n$ elements. According to the classical information theory, the $m$ i.i.d. random variables may have different probabilities to get different values, so the probability of sequence ${X}$ is equal to 
  \begin{equation}
	  P({X})=\prod_{j=1}^mp(\vec{r}_j)=\prod_{j=1}^mp_j^{r_j}(1-p_j)^{n-r_j}.
  \end{equation}
 Here, we still focus on the binary information sequence. Thus, $p_j$ is the probability of each unit in row $j$ to have value $1$. The $r_j$ is the number of units in row $j$ that have a value of $1$, and it will affect the process of signal generating when different constraints model it. 
 
 The value of $p_j$ still can be obtained by the average value of total units with value $1$ in each row as $p_j=r_j/n$. When the $j$th variable is represented by $b_j$, then the AEP will be generalized as
  \begin{equation}
	 \begin{aligned}
		 \frac{1}{n}\ln P({X})&=\sum_{j=1}^m[\frac{r_j}{n}\ln\frac{r_j}{n}+\frac{n-r_j}{n}\ln\frac{n-r_j}{n}]\\
		 &\rightarrow-\sum_{j=1}^ms(b_j)
	 \end{aligned}
  \end{equation}  
 Sequences belonging to the typical set of this system with multivariate independence information source should satisfy the following condition 
 \begin{equation}
	 T_{\epsilon}=\{{X}|e^{-n\sum_{j=1}^ms(b_j)-\epsilon}\leq P({X})\leq e^{-n\sum_{j=1}^ms(b_j)+\epsilon}\}.
 \end{equation}
 The space to store the information carried by it is equal to 
 \begin{equation}
	 \ln|T_{\epsilon}|=n\times \sum_{j=1}^ms(b_j).
 \end{equation}
 The limit of information storage is still decided by the uncertainty of the information source, even if there are $m$ independent variables. 
 The information sequences that are generated by the multivariate information source can still be described by the canonical and microcanonical ensemble. 
 Following, we will introduce the details about the ensemble description of the information sequence generated by the $m$ multivariate independent information sources.
 
 \subsection{Canonical ensemble description of information sequence from i.i.d multivariate source.}\label{Section_can_r_multi}
 The information sequence ${{X}}$ generated by the independent variables with different probability distribution can be modeled by the matrix ${X}$ based on the row local constraints $\vec{r}({X})=[r_1,r_2,\cdots,r_m]$, where $r_j=\sum_{i=1}^nx_{ji}$. The maximum likelihood parameter $\vec{\beta}^*=[\beta^*_1,\beta^*_2,\cdots,\beta^*_m]$ has $m$ elements~\cite{zhang2022strong}. 
 
 The \emph{Hamiltonian} is still the linear combination of the constraints and parameters $H({X})=\sum_{j=1}^m\beta^*_jr_j$. In the binary case, the partition function of this matrix ensemble is 
  \begin{equation}
	  Z(\vec{\beta}^*)=\prod_{j=1}^m(1+e^{-\beta^*_j})^n.
  \end{equation}
  Then we can have the canonical probability of the state $\mathbf{X}$ with the localized maximum likelihood parameters $\vec{\beta}^*$ as
  \begin{equation}
	 P_{\textrm{can}}({X}|\vec{\beta}^*)=\prod_{j=1}^m\frac{e^{-\beta^*_jr_j}}{(1+e^{-\beta^*_j})^n}.
  \end{equation}
 The parameter $\beta_j^*$ is decided by the average value of row local constraints $\langle r_j\rangle$.
 
 The space to store the information carried by the information sequences $\mathbf{X}$ still can be quantified by the AEP. 
 The rescaled logarithm of the probability is equal to 
  \begin{equation}
	  \frac{1}{n}\ln P_{\textrm{can}}({X}|\vec{\beta}^*)=\frac{1}{n}\sum_{j=1}^m\sum_{i=1}^n\ln\frac{e^{-\beta^*_jx_{ji}}}{1+e^{-\beta^*_j}}.
  \end{equation}
 When the length $n$ goes to infinite, the probability of the units in $j$th row to have value $1$ is equal to average value of $x_{ji}$ as
 \begin{equation}
	 \langle x_{ji}\rangle=\frac{e^{-\beta^*_j}}{1+e^{-\beta^*_j}}=\frac{\langle r_j\rangle}{n}.
 \end{equation}
 The sum of the logarithms of the probability for the $m\times n$ units will equal the sum of the $m$ variables Shannon entropy
 \begin{equation}
	 \begin{aligned}
		 \lim_{n\rightarrow\infty}\frac{1}{n}\ln P_{\textrm{can}}({X}|\vec{\beta}^*)&\rightarrow -\sum_{j=1}^m s(b_j).
		\end{aligned}
	\end{equation}	
	When the average value of $r_j$ equal to the hard constraints in the conjugated microcanonical ensemble as $\langle r_j\rangle=r^*_j$, the value of the Shannon entropy of the information sources with $m$ variables is
	\begin{equation}
		\begin{aligned}
			\sum_{j=1}^m s(b_j)=\sum_{j=1}^m[\frac{r^*_j}{n}\ln\frac{r^*_j}{n}+\frac{n-r^*_j}{n}\ln\frac{n-r^*_j}{n}].\\
	 \end{aligned}
 \end{equation}

 When the value of $\epsilon$ equals 0, the space to store the information carried by the sequences still contains in the typical set $T_{\textrm{can}}^{\epsilon=0}$, which satisfies the following condition 
 \begin{equation}
	 T_{\textrm{can}}^{\epsilon=0}=\{{X}|P_{\textrm{can}}({X}|\vec{\beta}^*)= e^{n\sum_{j=1}^m s(b_j)}\}.
 \end{equation}
 The logarithm of the number of sequences in the typical set is also equal to $\ln |T_{\textrm{can}}|=n\sum_{j=1}^m s(b_j)$, which is the Shannon entropy of the sequences $S_{\textrm{can}}({X})$
  \begin{equation}
	 \ln |T_{\textrm{can}}^{\epsilon=0}|=S_{\textrm{can}}({X})=\sum_{j=1}^m\ln[\frac{n^n}{{r^*_j}^{r^*_j}(n-r^*_j)^{n-r^*_j}}].
  \end{equation} 
 It is equal to the $\ln|T_{\epsilon}|=n\times \sum_{j=1}^ms(b_j)$ in the classical information theory. 
 
 This result shows that the method in the classical information theory is a particular case of the canonical ensemble description of information sequences. They are both the maximum entropy approximation of the real process of signal generation.

 \subsection{Microcanonical ensemble description of the information sequences from i.i.d multivariate source}
 When the multivariate information source is under hard constraints, i.e., the energy that each unit can use in the information source is finite, the signal generation process needs to be described by the microcanonical ensemble.
  
 The hard constraints are the total number of units with $1$ in each row equal to $r^*_j$. The constraints are the $\vec{r}^*=[r^*_1,r^*_2,\cdots,r^*_j,\cdots,r^*_m]$. The probability of each sequence appearing in the signal generation is equivalent, and it is decided the total number of configurations of the local row-constrained sequences with constraints $\vec{r^*}$ as 
  \begin{equation}
	  P_{\textrm{mic}}({X}|\vec{r}^*)=1/\prod_{j=1}^m\binom{n}{r^*_j}.
  \end{equation}
 According to the asymptotic equipartition property, all the sequences in the microcanonical ensemble belong to the typical set of it, the space to store the information carried by it is equal to the Boltzmann entropy of this sequence as
 \begin{equation}
	 \ln|T_{\textrm{mic}}^{(\vec{r}^*)}|=\sum_{j=1}^m\ln\binom{n}{r^*_j}=\ln \Omega_{\vec{r}^*}.
 \end{equation}
 The calculation of the total number of the configurations in the information sequences that are described by the microcanonical ensemble is 
 \begin{equation}
	\Omega_{\vec{r}^*}=\prod_{j=1}^{m}\frac{n!}{r^*_j!(n-r^*_j)!}\rightarrow\prod_{j=1}^{m}\frac{n^n}{{r^*_j}^{r^*_j}(n-r^*_j)^{(n-r^*_j)}}.
 \end{equation}
 When $n\rightarrow\infty$, the calculation requires the value of $r^*_j$ to grow like $O(n)$. Then, based on Stirling's formula, we can have the value of $\Omega_{\vec{r}^*}$.

 The space needed to store the information carried by the microcanonical ensemble described information sequence is 
 \begin{equation}
	\ln|T_{\textrm{mic}}^{(\vec{r}^*)}|=\ln \Omega_{\vec{r}^*}\stackrel{n\rightarrow\infty}{\rightarrow}\ln \prod_{j=1}^{m}\frac{n^n}{{r^*_j}^{r^*_j}(n-r^*_j)^{(n-r^*_j)}} .
\end{equation}

This microcanonical ensemble description of the information sequences clearly shows that the space needed to store the information carried by those sequences is limited by the uncertainty of the whole sequences. 

\subsection{Ensemble equivalence in the multivariate i.i.d. informaiton sources}
The canonical and microcanonical ensemble descriptions of the multivariate i.i.d. information sources are based on the row local constraints assumption~\cite{zhang2022strong}. 

The space needed to store the information carried by those ensemble-described information sequences is the macroscopic property of the information sequences, and they are equal to the Boltzmann entropy of the microcanonical ensemble and the Gibbs entropy of the canonical ensemble. The difference in the limit of information storage for the two different ensemble descriptions equals the difference in their entropy as
\begin{equation}
	\ln|T_{\textrm{can}}^{\epsilon=0}|-\ln|T_{\textrm{mic}}^{(\vec{r}^*)}|=\Delta S^{(\vec{r}^*)}_{\textrm{can}}.
\end{equation} 
The value of it is decided by the local row constraints $\vec{r}^*=[r^*_1,r^*_2,\cdots,r^*_j,\cdots,r^*_m]$ as 

\begin{equation}
	\begin{aligned}
		\Delta S^{(\vec{r}^*)}_{\textrm{can}}&=S(P_{\textrm{mic}}||P_{\textrm{can}})\\
		&=\sum_{{X}\in\mathbf{X}}P_{\textrm{mic}}({X}|\vec{r}^*)\ln\frac{P_{\textrm{mic}}({X}|\vec{r}^*)}{P_{\textrm{can}}({X}|\vec{\beta}^*)}\\
		&=\sum_{j=1}^{m}[\ln \frac{n^n}{{r^*_j}^{r^*_j}(n-r^*_j)^{(n-r^*_j)}}-\ln\binom{n}{r^*_j}]\\
		&=\sum_{j=1}^m[\frac{1}{2}\ln[2\pi r^*_j(1-\frac{r^*_j}{n})]]
	\end{aligned}
\end{equation}

The measure level ensemble equivalence is the probability distribution of the canonical ensemble states converging to the conjugate microcanonical canonical ensemble. The tendency of the convergence can be quantified by the resca1 relative entropy between the microcanonical ensemble and canonical ensemble.
The rescaled difference between the logarithm of the two ensemble's probability is close to the rescaled total correlations $S(P_{\textrm{mic}}||P_{\textrm{can}})/n$. The asymptotic behavior of $\frac{1}{n}\Delta S^{(\vec{r}^*)}_{\textrm{can}}$ is decided by $m$, in this work, $m=o(n)$, thus the limit of $\frac{1}{n}\Delta S^{(\vec{r}^*)}_{\textrm{can}}$ is equal to 0 as
 \begin{equation}
	\begin{aligned}
		\lim_{n\rightarrow\infty}\frac{1}{n}\Delta S^{(\vec{r}^*)}_{\textrm{can}}&=\lim_{n\rightarrow\infty}\frac{1}{2n}\sum_{j=1}^m[\ln[2\pi r^*_j(1-\frac{r^*_j}{n})]]\\
		&=0,		
	\end{aligned}
 \end{equation}
which means the signal generation by the classical independent multivariate information sources is also under ensemble equivalence. That is why we can still use the AEP to estimate information-theoretical bounds. 
 
 \section{conclusions and discussion}
The space we need to store the information is decided by the number of configurations of the sequences used to carry the information generated by the information source. 
The probability of those equiprobable sequences in the typical set equals  $P(A)=e^{n\times s(x)}$, as the entropy of the information sequence equals the length times the entropy of the information source, $S(A)=n\times s(x)$. Then, the limit of space needed to store the information generated by the information sources is decided by the uncertainty of the information source. Thus, the asymptotic equipartition property can be explained in two aspects: 1. Sequences belonging to the typical set carry almost all the information; 2. The Shannon entropy of the information source decides the probability of sequences in the typical.

The asymptotic equipartition property also has a physic version. In statistical physics, there is a consensus that the microcosmic behavior of those particles decides the macroscopic property of a thermodynamic system. This property is based on the fact that the number of particles in a macroscopic thermodynamic system is big enough (i.e., a system with $10^{23}$ particles is big enough). This big scale is analogous to the condition of asymptotic equipartition property, which requires the length of the information sequence to grow infinitely. Thus, the length of the information sequences going to infinite is equivalent to the thermodynamic limit in statistical physics. The typical set also exists in the systems under statistical ensemble descriptions. Generally, the canonical ensemble with the Boltzmann distribution describes systems with soft constraints. However, when the systems are under conjugate hard constraints, they need to be described by the microcanonical ensemble, which states are equiprobable. Obviously, the states in the conjugate microcanonical ensemble are part of the \textit{typical set} of the canonical ensemble. The question is if the conjugate microcanonical ensemble is the typical set of the canonical ensemble or if the two ensembles are equivalent. We find the typical set of the canonical ensemble described information sequences will converge to the microcanonical ensemble when the length of the information sequences goes infinite $T^{\epsilon}_{\textrm{can}}\overset{n\rightarrow\infty}{=}T^{\epsilon}_{\textrm{mic}}$ both for the classical information source and the multivariate independent information source, which means the classical information sequence and information sequences generated by the multivariate independent information source are all under ensemble equivalence. 

Therefore, the length of the information sequence goes to infinite is a thermodynamic limit. The typical set of states in the canonical ensemble description is the states of the conjugate microcanonical ensemble. The classical information theory is a special case of the canonical ensemble description for the information sequence. The effectiveness of the classical information theory is a representation of the ensemble equivalence, which is an isotope of the asymptotic equipartition property.

\section*{Acknowledgments}
This work is supported by the Scientific Research Funding of Jiangsu University of Science and Technology (No.1052932204), the National Natural Science Foundation of China (Grant No. 62303198), the Research Initiation Fund for Senior Talents of Jiangsu University (No. 5501170008).

\bibliography{qi_test_google.bib}
\appendix

\color{blue}

\end{document}